\documentclass[prl,twocolumn,showpacs,superscriptaddress,floatfix]{revtex4-1}
\usepackage{graphicx,amsfonts,amssymb,amsmath,xspace,dsfont}
\usepackage[colorlinks=true,citecolor=green,linkcolor=red,urlcolor=blue]{hyperref}

\usepackage{color}
\definecolor{g}{rgb}{.1,0.4,.1} 
\definecolor{b}{rgb}{0,0.2,1}
\definecolor{rouge}{rgb}{0.82,0.,0.}
\definecolor{vert}{rgb}{0.,0.82,0.}
\definecolor{orange}{rgb}{1,0.5,0.}
\definecolor{bleu}{rgb}{0.,0.,0.82}
\definecolor{m}{rgb}{0.82,0.,0.82}
\definecolor{vert2}{rgb}{0.,0.5,0.}
\definecolor{rougeclair}{rgb}{1.0,0.7,0.7}

\begin{document}

\title{Ising anyons with a string tension}

\author{Marc Daniel Schulz}
\email{schulz@fkt.physik.tu-dortmund.de}
\affiliation{Lehrstuhl f\"ur Theoretische Physik I, Technische Universit\"at Dortmund, Otto-Hahn-Stra\ss e 4, 44221 Dortmund, Germany}
\affiliation{Laboratoire de Physique Th\'eorique de la Mati\`ere Condens\'ee,
CNRS UMR 7600, Universit\'e Pierre et Marie Curie, 4 Place Jussieu, 75252
Paris Cedex 05, France}

\author{S\'{e}bastien Dusuel}
\email{sdusuel@gmail.com}
\affiliation{Lyc\'ee Saint-Louis, 44 Boulevard Saint-Michel, 75006 Paris, France}

\author{Gr\'{e}goire Misguich}
\email{gregoire.misguich@cea.fr}
\affiliation{Institut de Physique Th\'eorique, CNRS URA 2306, CEA, 91191 Gif-sur-Yvette, France}

\author{Kai Phillip Schmidt}
\email{schmidt@fkt.physik.tu-dortmund.de}
\affiliation{Lehrstuhl f\"ur Theoretische Physik I, Technische Universit\"at Dortmund, Otto-Hahn-Stra\ss e 4, 44221 Dortmund, Germany}

\author{Julien Vidal}
\email{vidal@lptmc.jussieu.fr}
\affiliation{Laboratoire de Physique Th\'eorique de la Mati\`ere Condens\'ee,
CNRS UMR 7600, Universit\'e Pierre et Marie Curie, 4 Place Jussieu, 75252
Paris Cedex 05, France}

\begin{abstract}

We consider the string-net model on the honeycomb lattice for Ising anyons in the presence of a string tension. This competing term induces a nontrivial dynamics of the non-Abelian anyonic quasiparticles and may lead to a breakdown of the topological phase. Using high-order series expansions and exact diagonalizations, we determine the robustness of this doubled Ising phase which is found to be separated from two gapped phases.  An effective quantum dimer model emerges in the large tension limit giving rise to two different translation symmetry-broken phases. Consequently, we obtain four transition points, two of which are associated with first-order transitions whereas the two others are found to be continuous and provide examples of recently proposed Bose condensation for anyons. 

\end{abstract}


\maketitle

More than 20 years after its discovery \cite{Wen89_1,Wen89_2,Wen90_1}, topological quantum order remains one of the most fascinating fields of condensed matter physics. Topologically ordered systems are characterized by several features such as, e.g., the topological degeneracy, exotic braiding statistics, or  long-range entanglement (see Ref. \cite{Wen13} for a recent review). Contrary to conventional phases, topological phases cannot be described by a local order parameter so that the Landau-Ginzburg theory cannot be used to investigate transitions between them. In this context, new tools have been developed to understand transition mechanisms. Among them, an appealing approach relying on condensation of bosonic quasiparticles \cite{Bais09_1} has been proposed to determine some possible connections between different phases. We refer the interested readers to  Refs.~\cite{Bais09_2,Barkeshli10,Burnell11_2,Burnell12,Burnell13} for concrete examples in lattice models  and to Refs.~\cite{Kong13,Fuchs13,Fuchs14,Eliens13} for more mathematical considerations.  However, a complete description of topological phase transitions is still missing and, in particular, a classification of universality classes for the critical properties is still to be established. From that respect, it seems essential to study microscopic models in order to explore possible scenarios. 

In this Rapid Communication, we analyze the zero-temperature phase diagram of the string-net model \cite{Levin05} defined on the honeycomb lattice with Ising anyons in the presence of a string tension. First, we give some properties of the unperturbed Ising string-net model and we discuss several limiting cases allowing for a qualitative understanding of the phase diagram. To go beyond, we compute high-order series expansions of the low-energy spectrum in two limiting cases that we compare with exact diagonalization (ED) results. Apart from a trivial (polarized) phase and the doubled Ising (DIsing) topological phase, we find two different translation symmetry-broken phases emerging from an effective quantum dimer model whose analysis is given in Ref.~\cite{Supp_Mat}.  Furthermore, our results also suggest the possibility of universality classes associated with the condensation of Ising quasiparticles. 

%
%
{\em Hilbert space}.
%
%
Microscopic degrees of freedom of the string-net model are defined on the edges of a trivalent graph \cite{Levin05}. For the Ising theory considered thereafter, they can be in three different states $|1\rangle$, $|\sigma \rangle$, and $|\psi \rangle$. The Hilbert space ${\mathcal H}$ is then defined by the set of states that satisfy the so-called branching rules (at each vertex) stemming from the  SU$(2)_2$ fusion rules
%
%
\begin{eqnarray}
1 \times a = a \times 1&=& a, \:\: \forall a \in \{1,\sigma,\psi\}, \label{eq:fusion1} \\
\sigma \times \sigma= 1+\psi, \:\: \sigma \times \psi&=& \psi \times \sigma=\sigma, \:\: \psi\times \psi= 1.
\label{eq:fusion2}
\end{eqnarray} 
%
%
For any trivalent graph with $N_{\rm v}$  vertices, the dimension of the Hilbert space is then given by \cite{Gils09_3}
%
%
\begin{equation}
\label{eq:dimH}
\dim \mathcal{H}= 2^{N_{\mathrm v}+1}+2^{N_{\mathrm v}/2}.
\end{equation} 
%
%

%
%
{\em Model}.
%
%
Let us consider the following Hamiltonian
%
%
\begin{equation}
 \label{eq:ham}
H=- J_{\rm p} \sum_p  \delta_{\Phi(p),1} - J_{\rm e} \sum_e \delta_{l(e),1},
\end{equation} 
%
%
first introduced in Refs.~\cite{Gils09_1,Gils09_3}. The first term is  the string-net Hamiltonian introduced by Levin and Wen \cite{Levin05}. It involves the projector $\delta_{\Phi(p),1}$ onto  states with no flux $\Phi(p)$ through \mbox{plaquette $p$}. The second term is diagonal in the canonical basis introduced above since $\delta_{l(e),1}$ is the projector onto state $|1\rangle$ on edge $e$. This latter term is a string tension since it breaks the topological properties of the ground state described in Ref.~\cite{Levin05}. Without loss of generality, we set $J_{\rm p}=\cos \theta$ and $J_{\rm e}=\sin \theta$.

For $\theta=0$, the system is, by construction, in a doubled (achiral) Ising topological phase, dubbed DIsing in the following \cite{Burnell12,Liu14}. Consequently, the degeneracy of the eigenstates depends on the graph topology.  For instance,  the degeneracy of the $k^{\rm th}$ energy level (\mbox{$E_k=-N_{\rm p}+k$}) on a torus with $N_{\rm p}$ plaquettes  is  \cite{Dusuel14}
%
%
\begin{equation}
\label{eq:deg_Fibo}
\mathcal{D}_k= 
\left(
\begin{array}{c}
N_{\mathrm p}
\\
k
\end{array}
\right) \big[1+6(-1)^k+2 \times 3^k\big],
\end{equation} 
%
%
where the binomial coefficient  results from the different ways to choose $k$ plaquettes carrying the flux excitations among $N_{\mathrm p}$. In particular, one finds $\mathcal{D}_0=9$ ground states  that are labeled by the (trivalued) flux contained in each of the two non-contractible loops of the torus \cite{Burnell12}. Note that, using the Euler-Poincar\'e relation for a trivalent graph on this genus-one surface ($N_{\mathrm v}=2 N_{\mathrm p}$), it is easy to check that $\dim \mathcal{H}=\sum_{k=0}^{N_{\rm p}} \mathcal{D}_k$ matches with Eq.~(\ref{eq:dimH}).  

One goal of the present work is to analyze the stability of this topological phase when  $J_{\rm e}$ is switched on as well as to characterize the transition between various phases. Indeed, for $\theta=\pi/2$, the (unique) ground state is the state where all edges are in the state $|1\rangle$. Thus, there must be at least one phase transition  in the range $[0,\pi/2]$. For $\theta=3\pi/2$, the ground state is infinitely many degenerate in the thermodynamical limit so that (at least) one phase transition is expected in the range $[3\pi/2,2\pi]$. Finally, for $\theta=\pi$, the ground-state degeneracy (on a torus) is given by $\mathcal{D}_{N_{\mathrm p}}$ so that transitions must also occur in the range $[\pi/2,3\pi/2]$.
In the following, we consider the simplest two-dimensional trivalent graph, namely, the honeycomb lattice.\\

%
%
{\em The ``simple" case: $\theta\in [0,\pi]$}.
%
%
 To determine  the boundaries of the DIsing topological phase (\mbox{around $\theta=0$}), we computed the ground-state energy as well as the quasiparticle gaps by means of high-order series expansions in powers of $J_{\rm e}/J_{\rm p}$ (lengthy expressions are given in Ref.~\cite{Supp_Mat}) using various techniques \cite{Loewdin62,Takahashi77,Knetter00}. 
In the vicinity of $\theta=0$, one must make the distinction between two different low-energy gaps corresponding to quasi-$\sigma$ and quasi-$\psi$ excitations. These excitations are usually referred to as $\sigma_L \sigma_R$ and $\psi_L \psi_R$ in the literature (see for instance Ref.~\cite{Burnell12}) but, for simplicity, we will adopt here a quasiparticle language keeping in mind that these are achiral objects. Contrary to quasi-$\psi$ excitations,  a single quasi-$\sigma$ excitation cannot exist on a compact surface such as the torus because of the branching rules \cite{Dusuel14}. Using standard extrapolation methods, we determined the points where these gaps vanish and thus established the stability range of the DIsing phase. However, if  level crossings due to higher-energy levels are present, a first-order transition may also arise and cannot be captured by our perturbative approach that only deals with low-energy states. To check the validity of the conclusions drawn from the series expansions, we 
performed ED of $H$ using periodic boundary conditions and systems with unit vectors of equal norms forming an angle of $\pi/3$.  On the torus, the Hamiltonian $H$ can be split into topological different sectors that must not be confused with the nine flux sectors discussed previously for $\theta=0$. Indeed, for any $\theta$, branching rules impose that a  $|\sigma \rangle$ link is always connected to a single $|\sigma \rangle$ link. In particular, there exist non-contractible loops of $|\sigma \rangle$ links enclosing the torus. Fusion rules impose that $H$ only conserves the parity of the number of such loops and, since there are two independent non-contractible loops on the torus, one has $2\times2$ different sectors for any $\theta$. 

\begin{figure}[t]
\includegraphics[width= 0.8\columnwidth]{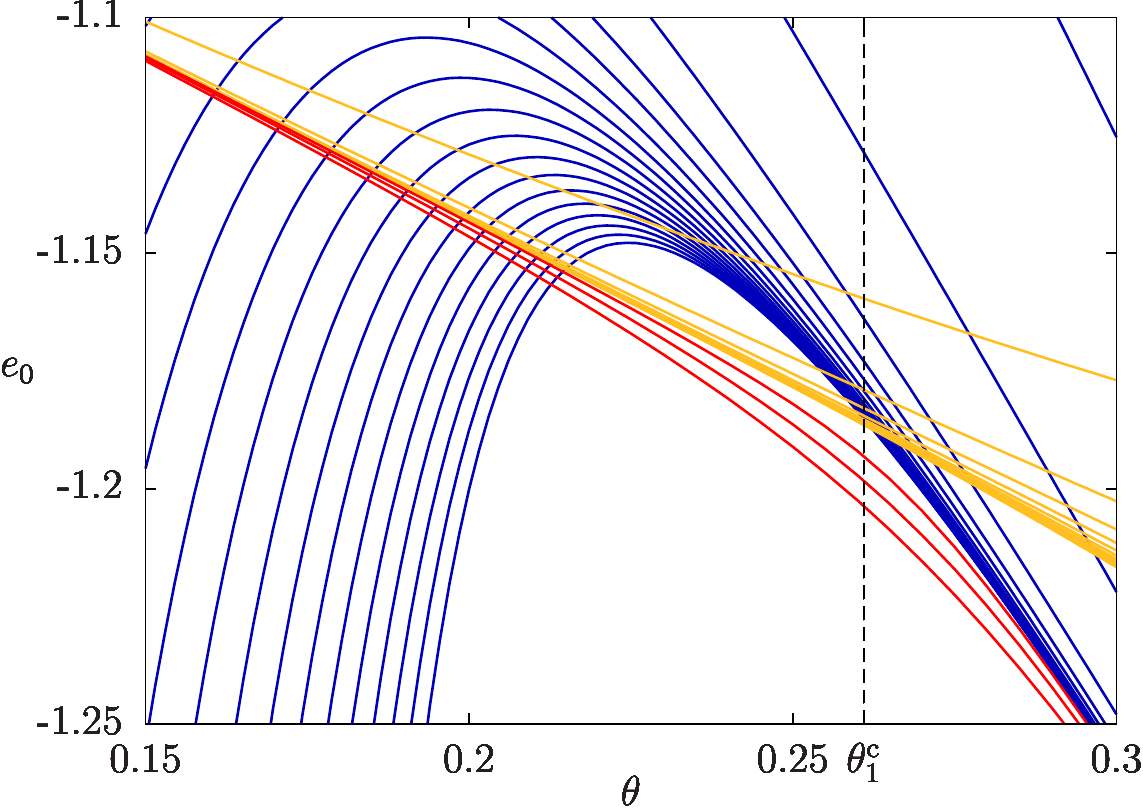}
\caption{(Color online)
From top to bottom, ground-state energy per plaquette $e_0$ computed from series expansions at order 1-11 around $\theta=0$ (yellow) and at order 1-18 around $\theta=\pi/2$ (blue). For comparison, ED results (red) are also displayed for $N_{\mathrm p}=7, 9,13$ (from bottom to top).}
\label{fig:gse}
\end{figure}

\begin{figure}[t]
\includegraphics[width= 0.8\columnwidth]{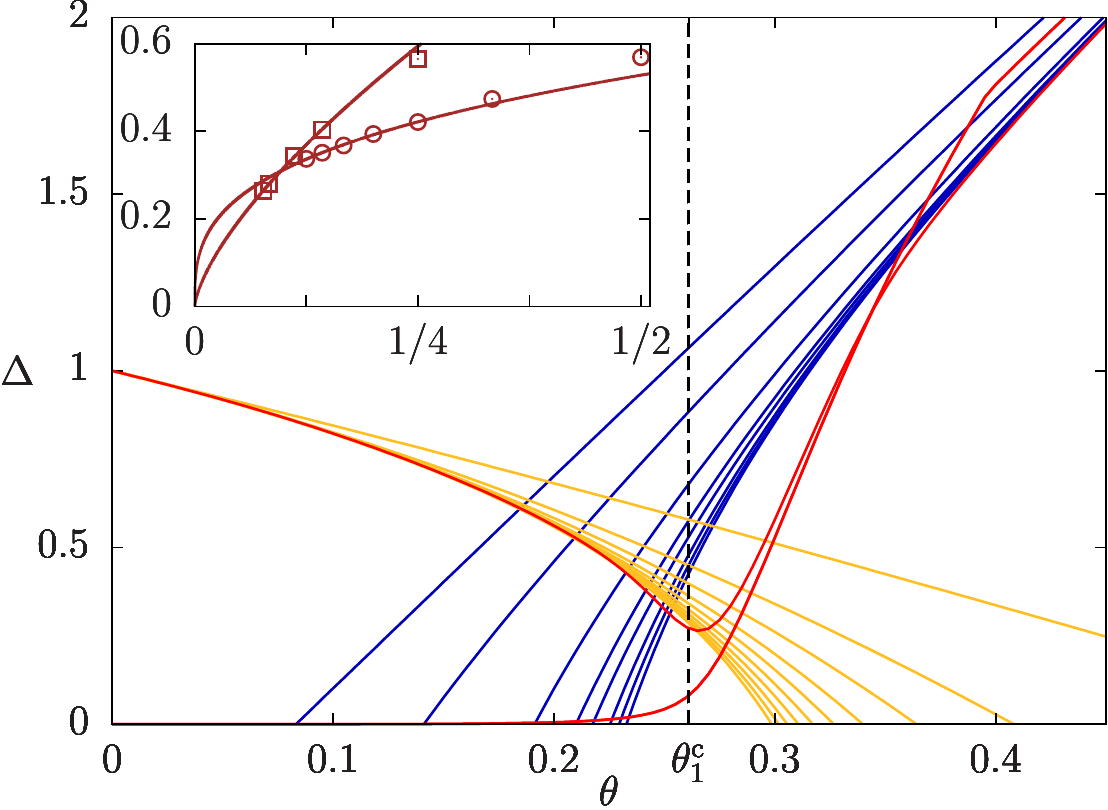}
\caption{(Color online) 
From top to bottom, low-energy gap $\Delta$ computed from series expansions at order 1-10  around $\theta=0$ (yellow) and at order 1-8 around $\theta=\pi/2$ (blue). For comparison, the first nine excitation energies obtained from ED (red) are shown for $N_{\mathrm p}=13$. The first excited level is eightfold degenerate. Inset: Minimum of ninth excitation energy as a function of  $N_{\mathrm p}^{-1}$ computed from ED (squares); minimum of $\Delta$ as a function of $n^{-1}$ (circles) computed at the crossing point between order $n$ series performed around $\theta=0$ and $\theta=\pi/2$. Lines are power law fits consistent with  $\displaystyle{\lim_{n,N_{\mathrm p} \to 0}} \Delta =0$.}
\label{fig:gap1}
\end{figure}

We display in Fig.~\ref{fig:gse} a comparison between the ED results and the series expansions for the ground-state energy performed around $\theta=0$ (red) and $\theta=\pi/2$ (blue). As can be seen, at each order, the series intersect at two different points. Similarly, we show in Fig.~\ref{fig:gap1} the results for the low-energy gap that intersect in a unique point. 
After extrapolations, using the same analysis as in Ref.~\cite{Schulz13}, we found that all these crossing points converge towards a unique value defining a second-order transition point at $\theta^{\rm c}_1 \simeq 0.261$. This point also matches with the position of the infinite-size extrapolation of the gap minimum as well as the minimum of $\partial^2_\theta e_0$ computed from ED. This critical point separates the DIsing phase originating from $\theta=0$ from the polarized (non-topological) phase near $\theta=\pi/2$. 
%
%
%
%

We stress that the relevant low-energy gap in the ED must be interpreted as a single quasiparticle gap associated with quasi-$\psi$ excitations. The corresponding energy level is indeed adiabatically connected to $E_1=E_0+1$ at $\theta=0$. Within the topological symmetry-breaking formalism proposed in Ref.~\cite{Bais09_1}, if quasi-$\psi$ excitations condense while quasi-$\sigma$ excitations remain gapped, one should switch towards another topological phase, which is not the case here. In this framework, the only possibility to enter a non-topological phase is that the quasi-$\sigma$ gap also vanishes at $\theta^{\rm c}_1$. As explained above, this gap cannot be observed in ED on a torus but it can be computed perturbatively since this approach is independent of the surface topology in the thermodynamical limit. Series expansions of both quasiparticle gaps (quasi-$\psi$ and quasi-$\sigma$) are given in Ref.~\cite{Supp_Mat}. As can be checked explicitly, they are strictly identical up to order $4$ and differ beyond. However, the sign of the 
(tiny) difference between both gaps changes at each order, which is compatible with a simultaneous vanishing of these gaps at $\theta^{\rm c}_1$. Note that we performed similar calculations for the ladder geometry and we found that both gaps are identical at all orders we \mbox{computed \cite{Schulz14}}. 
For $\theta \in[\pi/2, \pi[$, we did not find any indication of a transition but, as for the Fibonacci theory \cite{Schulz13}, the first derivative of the ground-state energy per plaquette $\partial_\theta e_0$ displays a jump at $\theta=\pi$ indicating a first-order transition. Thus, the trivial phase originating from $\theta=\pi/2$ extends from $\theta^{\rm c}_1$ (second-order transition point) to $\pi$ (first-order transition point).\\

%
%
{\em The ``original" case: $\theta\in [3\pi/2,2\pi]$}.
%
%
For $\theta=3\pi/2$, the ground-state manifold is spanned by all states minimizing the number of edges in state $|1\rangle$. Interestingly, for the Ising theory, fusion (branching) rules allow some states without any $|1\rangle$ bond provided each vertex touches exactly one $|\psi\rangle$ and two $|\sigma\rangle$. These constraints are nothing but those of hard-core dimer coverings of the hexagonal lattice if the state $|\psi\rangle$ is viewed as a bond occupied by a ''dimer." 
%
%
%
%

%
%
\begin{figure}[t]
\includegraphics[width= 0.8\columnwidth]{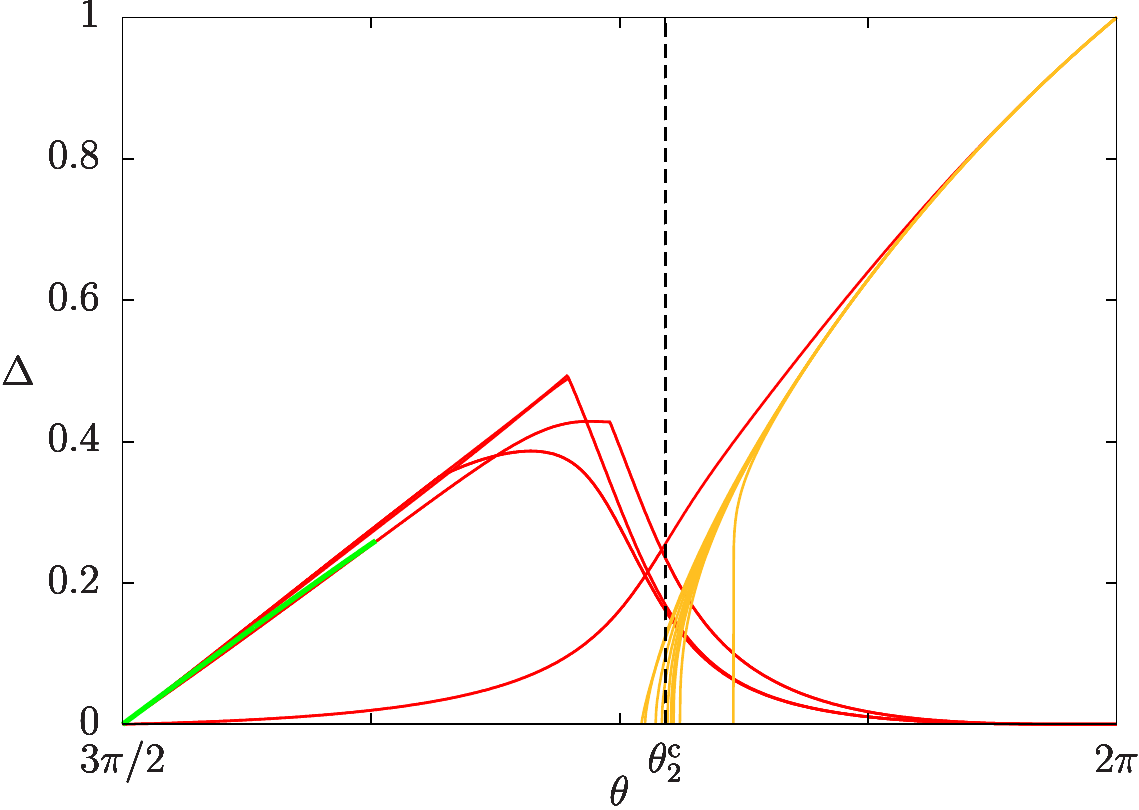}
\caption{(Color online)
Low-energy gap $\Delta$ computed from ressummed series expansions around $\theta=0$ [{\it d}logPad\'e approximants $[m,n]$ for $m,n \geqslant 3$ are displayed (yellow)] and from the expression (\ref{eq:gap_eff}), which is a good approximation of the gap around $\theta=3\pi/2^+$ (green). 
The first nine excitation energies obtained from ED (red) are shown for  $N_{\mathrm p}=12$.}
\label{fig:gap2}
\end{figure}
%
%

The exponential ground-state degeneracy at $\theta=3\pi/2$ prevents a simple series expansion around this point. In addition, the alternating signs in the series around $\theta=0^{-}$ prevents from an analysis similar to the one used in the range $[0,\pi/2]$. However, a close inspection of the gap series expansion around $\theta=0$ indicates a transition point near $\theta^{\rm c}_2\simeq 5.57$.  As can be seen in Fig.~\ref{fig:gap2},  ED results are consistent with a unique phase transition in the range $[3\pi/2,2\pi]$, but accessible sizes are definitely too small to characterize properly the phase for $\theta \in [3\pi/2,\theta^{\rm c}_2[$ as well as the nature of the transition. To gain a deeper understanding of this region, we derived the low-energy effective theory near $\theta=3\pi/2$, at leading order,  by considering the effect of the string-net Hamiltonian on the infinitely many degenerate ground-state manifold of the unperturbed ($J_{\rm p}=0$) problem. The effective Hamiltonian can be written in the following form 
%
%
\begin{eqnarray}
H_{\rm eff}= -\frac{J_{\rm p}}{4} \sum_p \:\:  \Big[&t& 
\Big(
\left|\hspace{-1mm}
\begin{array}{c}\includegraphics[width=0.5cm]{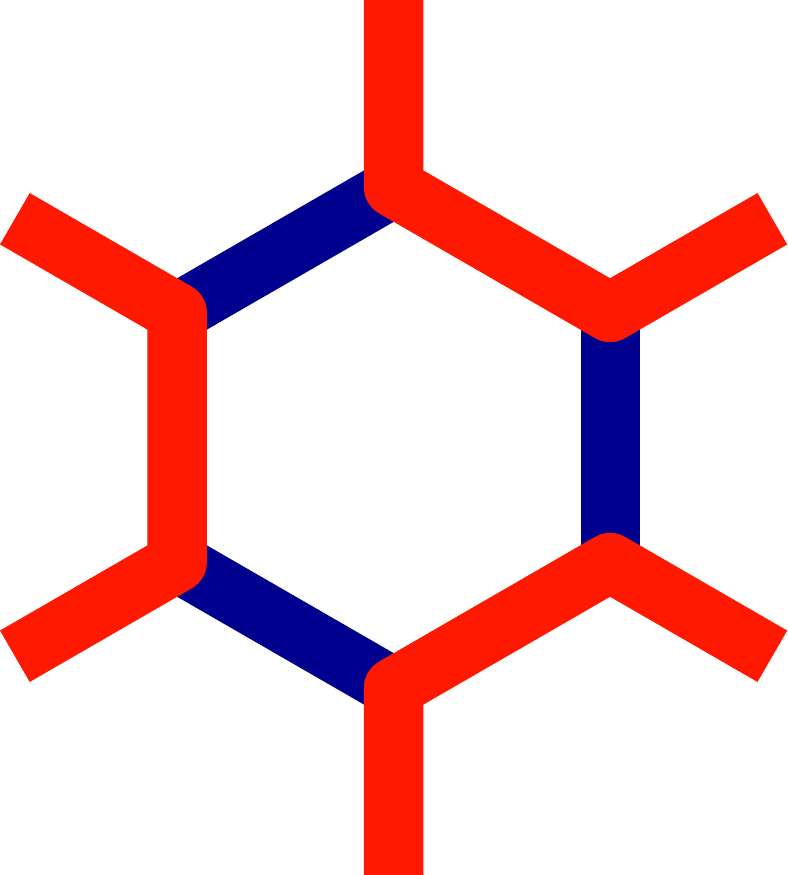}\end{array} 
\hspace{-1mm}\right\rangle 
\left\langle\hspace{-1mm}
\begin{array}{c}\includegraphics[width=0.5cm]{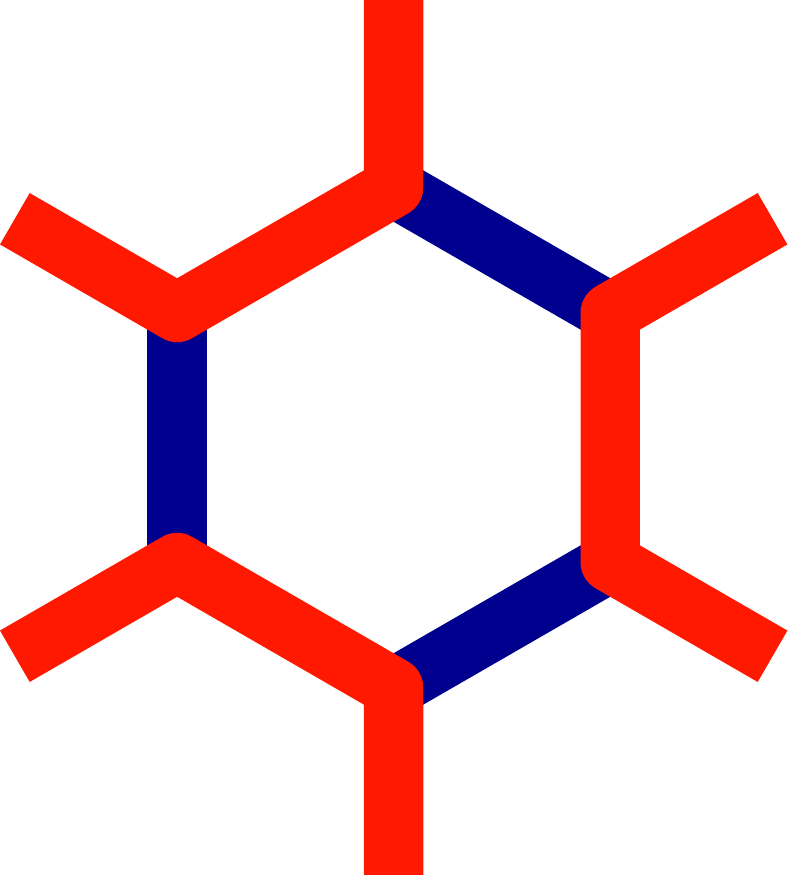}\end{array}
\hspace{-1mm}\right|
+
\left|\hspace{-1mm}
\begin{array}{c}\includegraphics[width=0.5cm]{./figures/flip_plaq2.pdf}\end{array}
\hspace{-1mm}\right\rangle
\left\langle\hspace{-1mm} 
\begin{array}{c}\includegraphics[width=0.5cm]{./figures/flip_plaq1.pdf}\end{array}
\hspace{-1mm}\right|
\Big)
+ \nonumber \\
&v&\:\: 
\left|\hspace{-1mm} 
\begin{array}{c}\includegraphics[width=0.5cm]{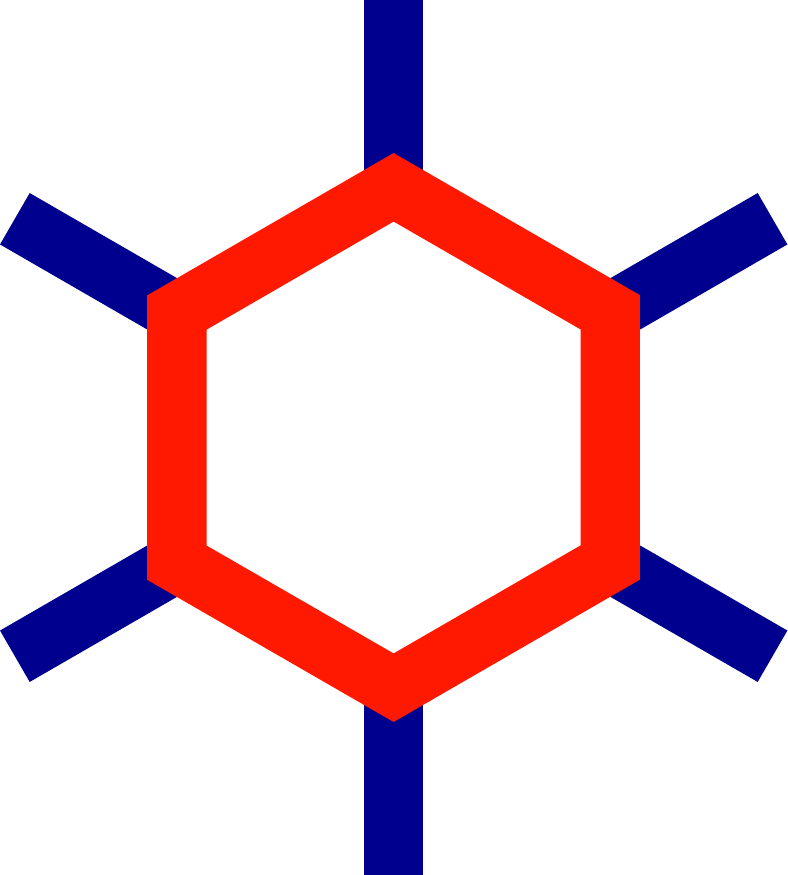}\end{array}
\hspace{-1mm}\right\rangle 
\left\langle\hspace{-1mm} 
\begin{array}{c}\includegraphics[width=0.5cm]{./figures/diag_plaq.pdf}\end{array}
\hspace{-1mm}\right|
 + \mathds{1} \Big],
\label{eq:hameff}
\end{eqnarray}
%
%
with $t=1/2$ and $v=1$. In this representation, a red (blue) link corresponds to a $|\sigma\rangle$ ($|\psi\rangle$) state. We emphasize that $H_{\rm eff}$ is only valid to compute the correction of order 1 ($\propto J_{\rm p}$) to the spectrum of $H$ near $\theta=3\pi/2$. This effective Hamiltonian consists of two terms: a kinetic term $t$ acting on ``flippable" plaquettes and a potential term $v$ proportional to the number of empty plaquettes (without dimers). This model looks very similar to the famous quantum dimer model introduced by Rokhsar and Kivelson on the square lattice \cite{Rokhsar88} and later studied on the honeycomb lattice \cite{Moessner01}.  The only difference between both models comes from the potential term that, in the Rokhsar-Kivelson model, is proportional to the number of flippable plaquettes.  
To our knowledge, the Hamiltonian (\ref{eq:hameff}) has yet to be studied and cannot be solved exactly for arbitrary couplings. However, for $J_{\rm p}>0$, it is possible to infer its low-energy properties by considering the limit $t/v \ll 1$ while keeping in mind that $t/v=1/2$ in our problem. For $t=0$ (and $J_{\rm p}>0$), the energy is minimized by maximizing the number of empty plaquettes (without dimers). In the thermodynamical limit, there are three possible ground states satisfying this constraint so that the system is in the so-called $\sqrt{3}\times\sqrt{3}$  star crystal (SC) phase \cite{Remark_SC} (see Ref.~\cite{Supp_Mat} for an illustration). First excited states are obtained by flipping one plaquette in one of these ground states.

The kinetic term $t$ induces quantum fluctuations that can be captured using perturbation theory.
At order 2 in $t/v$, the ground-state energy per plaquette is given by
%
%
\begin{equation}
\label{eq:gse_eff}
e_0=\frac{J_{\rm p}}{4} \left(-1-\frac{v}{3}-\frac{2}{9}\frac{t^2}{v}\right),
\end{equation} 
%
%
whereas the low-energy gap reads
%
%
\begin{equation}
\label{eq:gap_eff}
\Delta=\frac{J_{\rm p}}{4} \left(3v-\frac{4}{3} \frac{t^2}{v}\right).
\end{equation} 
%
%
Of course, one could reach higher orders in this $t/v$ expansions but it is not of crucial importance for the present study. Indeed, these expressions already provide a very good approximation of $e_0$  and $\Delta$ for $t/v=1/2$ since, as shown in Ref.~\cite{Supp_Mat},  they only differ from the infinite-size values extrapolated from ED results by less than $0.3\%$ and $1.5\%$, respectively.  In the range $[3\pi/2,\theta^{\rm c}_2[$, we thus find a gapped translation symmetry-broken phase with a three-fold degenerate ground state. Furthermore, the three momenta (center and the two corners of the hexagonal Brillouin zone) of the ground states in the SC coincide with the locations of the minima of the excitation gap in the DIsing phase. This suggests that the transition from the topological phase to the crystal also corresponds to a simultaneous condensation of the anyonic quasiparticles.

Furthermore, momenta of these SCs also minimize the dispersion in the range $\theta \in[0,\theta^{\rm c}_2[$ which is compatible with a second-order transition. 
This is an example of a continuous phase transition between a non-Abelian topological phase and a non topological (translation) symmetry-broken phase.\\

%
%
{\em The ``tricky" case: $\theta\in [\pi,3\pi/2]$}.
%
%
Obviously, the effective model (\ref{eq:hameff}) is also valid for $\theta=3\pi/2^-$ but, contrary to the case $J_{\rm p}>0$ where a SC is favored (see discussion above), the ground state is infinitely-many degenerate at $t=0$ so that it is difficult to consider a perturbative $t/v$ expansion. Consequently, we performed  ED of $H_{\rm eff}$ up to relatively large system sizes ($N_{\rm p}=63$) and we found that the point $t=1/2$, $v=1$ lies in the same phase as the point $t=1/2$, $v=0$ (see Ref.~\cite{Supp_Mat}). As discussed in Ref.~\cite{Moessner01} for $v=0$, the ground state displays the so-called plaquette order that breaks the translational symmetry. Thus, for $\theta=3\pi/2^-$,  we expect a non topological ordered (gapped) plaquette crystal (PC) phase with a three-fold degenerate ground state. However, in the absence of perturbative analysis near $\theta=\pi$ and because of important variations with the system size of the ED results (at least up to $N_{\rm p}=13$ which is our current limit), we did not succeed in  characterizing the whole interval $[\pi,3\pi/2]$.  However, we observe that $\partial_\theta e_0$ displays a jump for $\theta=\pi$ and $\theta=3\pi/2$ so that first-order transitions occur at these points.
%
%
\begin{figure}[t]
\includegraphics[width= 0.8\columnwidth]{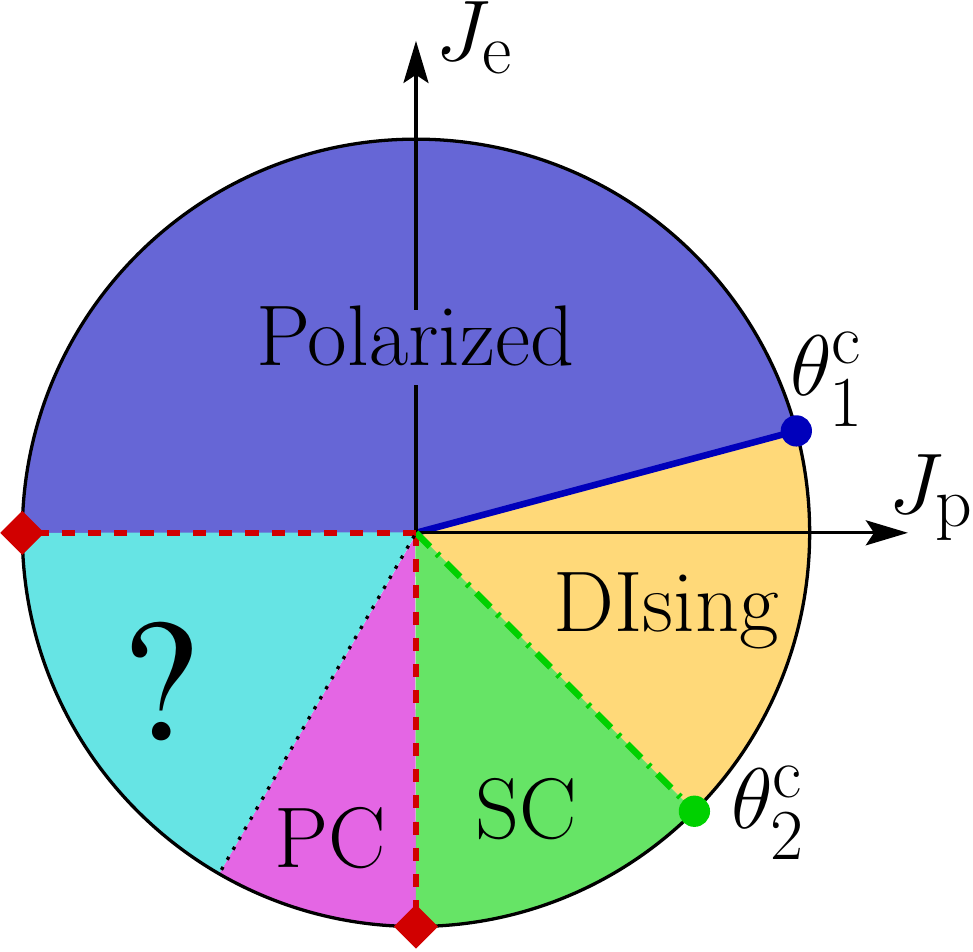}
\caption{(Color online) Sketch of the phase diagram. Four different gapped phases are identified among which  one is topologically ordered (DIsing) and two break the translation symmetry (SC and PC). Diamonds (circles) indicate first-order (second-order) transition points (see text for more details).}
\label{fig:diagram}
\end{figure}
%
%

%
%
{\em Summary and outlook}.
%
%
A sketch of the phase diagram gathering all informations discussed throughout this Rapid Communication is given in Fig.~\ref{fig:diagram}. One of the main results is the possibility to condense simultaneously quasi-$\psi$ and quasi-$\sigma$ excitations at the critical point $\theta_1^{\rm c}$, unveiling a likely new universality class. Unfortunately, in the absence of an alternative description, it is difficult to predict the associated critical exponents.  Setting $\theta_1^{\rm c}=0.261$, a standard ED data collapse analysis gives results that obey the hyperscaling relation for $z=1$ and $\nu\simeq 0.39$, and the resulting specific-heat exponent is $\alpha=2-\nu(2+z)\simeq 0.83$. One can also compute directly the exponent $z \nu$ using different {\it d}log Pad\'e approximants of the gap series and we found values in the range $[0.35,0.5]$. This rather broad range clearly indicates the lack of precision of such an approach for order 10 series, but suggests that this phase transition may belong to a new universality class. We hope 
that the present work will stimulate further studies to deepen our understanding of the topological phase transitions.\\


M.D.S. acknowledges financial support from the DFH-UFA. K.P.S. acknowledges ESF and EuroHorcs for funding through his EURYI. G.M. is supported by a JCJC grant of the Agence Nationale pour la Recherche (Project No. ANR-12-JS04-0010-01).


\begin{thebibliography}{28}
\bibitem [{Rem()}]{Remark_SC}%
  \BibitemOpen
  \href@noop {} {}\bibinfo {note} {In Ref.~\cite{Moessner01}, the star crystal
  is named ``columnar'' in analogy with the square lattice case for which this
  crystal indeed consists of dimers arranged in columns. In the hexagonal
  lattice, this ``columnar" crystal consists of triangular lattice of dimerless
  plaquettes with surrounded by six dimers so that we refer to it as the star
  crystal.}\BibitemShut {Stop}%
\end{thebibliography}

%


\onecolumngrid
\newpage

\section{\large Supplemental Material}

\section{Series expansions}

In the following, we give the series expansions in the different phases for the ground-state energy per plaquette $e_0$ and the quasiparticle gaps $\Delta^{\pm}$, for positive and negative signs of the dimensionless parameter $t=J_{\rm e}/J_{\rm p}=\tan \theta$ respectively. For the sake of clarity, instead of fractions, we give below the numerical values of the coefficients with 16 digits.

\subsection{Expansions in the vicinity \texorpdfstring{$\theta=0$}{theta=0}}

For $\theta=0$, the ground state is a nontrivial topological state without any flux in the plaquettes and elementary excitations are obtained by adding fluxes in plaquettes. These elementary excitations can be either $\sigma$-flux or $\psi$-flux although, on a torus, only single $\psi$-flux are allowed because of the fusion rules. Thus,  at order 0, the ground-state energy per plaquette is $e_0=-J_{\rm{p}}$ and the elementary gap is $\Delta^{\pm}_{\sigma,\psi}=J_{\rm{p}}$.

In the limit $|J_{\rm e}/J_{\rm p}| \ll 1$, we computed $e_0$ up to order $11$ using operator perturbation theory \cite{Takahashi77}, whereas  $\Delta^{\pm}_{\sigma,\psi}$ were obtained up to order $10$ using perturbative continuous unitary transformations \cite{Knetter00}. 
%
\begin{align}
e_0/J_{\rm{p}}=& -1-0.75 \,t-0.28125 \,t^2 -0.2109375 \,t^3-0.32958984375 \,t^4-0.5911865234375 \,t^5-1.24871826171875\, t^6 
\nonumber\\& -2.877640989091661\, t^7-7.114238025965514\, t^8 -18.48834346206744\, t^9-49.97040001522125\, t^{10}
\nonumber\\& -139.2884041430441\, t^{11},\nonumber
\displaybreak[0]\\
\nonumber\\
 \Delta^+_\sigma / J_{\rm{p}}=& \, 1-1.5\, t-1.875\, t^2-2.8125\, t^3-7.09375\, t^4 -16.52335611979167\, t^5-48.05825297037760\, t^6
 \nonumber\\& -133.1833968692356\, t^7-409.2101262765166\, t^8-1231.461932247098\, t^9-3900.735208706145\, t^{10},\nonumber
 \displaybreak[0]\\
 \nonumber\\
\Delta^+_\psi / J_{\rm{p}}=& \, 1-1.5\, t-1.875\, t^2-2.8125\, t^3-7.09375\, t^4-16.523681640625\, t^5-48.06215243869358\, t^6
\nonumber\\& -133.1707469092475\, t^7-409.2326163570086\, t^8-1231.410148056503\, t^9-3900.790315182440\, t^{10},\nonumber
\displaybreak[0]\\
\nonumber\\
\Delta^-_\sigma / J_{\rm{p}}=& \, 1+0.75\, t+0.09375\, t^2 +0.421875\, t^3+0.36962890625\, t^4+0.6806233723958333\, t^5+1.173678080240885\, t^6\nonumber
\nonumber\\&
+2.567433410220676\, t^7+5.445718276795046\, t^8+12.50392429651884\, t^9+29.51492451663127\, t^{10},\nonumber
\displaybreak[0]\\
\nonumber\\
\Delta^-_\psi / J_{\rm{p}}=&\, 1+0.75\, t+0.09375\, t^2 +0.421875\, t^3+0.36962890625\, t^4+0.6807861328125\, t^5+1.169758266872830 \, t^6\nonumber
\nonumber\\&
+2.556954012976752\, t^7+5.404811399512821\, t^8 +12.36748536516701\, t^9+29.04858138563500\, t^{10}.\nonumber
\end{align}
%

\subsection{Expansions in the vicinity \texorpdfstring{$\theta=\pi/2$}{theta=\pi/2}}

For $\theta=\pi/2$, the ground state is the unique polarized state made of $|1\rangle$-links and the elementary excitation consists in loops of six $|\sigma\rangle$-links or six $|\psi \rangle$-links around one plaquette. Thus,  at order 0, the ground-state energy per plaquette is $e_0=-3 J_{\rm{e}}$ and the gap $\Delta^\pm=6 J_{\rm{e}}$.

In the limit $|J_{\rm p}/J_{\rm e}| \ll 1$, we computed  $e_0$  up to order $18$ using a partitioning technique provided by L\"owdin \cite{Loewdin62} and the low-energy gap $\Delta^\pm$ were obtained up to order $8$ using operator perturbation theory \cite{Takahashi77}.

\begin{align}
e_0/J_{\rm{e}}=& -3 -0.25\, t^{-1}-0.03125\, t^{-2}-2.604166666666667\cdot 10^{-3}\, t^{-3}-1.785432449494949\cdot 10^{-4}\, t^{-4}
\nonumber\\&
-2.003492548591980\cdot 10^{-5}\, t^{-5}-3.518656736991765\cdot 10^{-6}\, t^{-6}-6.002642402965908\cdot 10^{-7}\, t^{-7}
\nonumber\\&
-1.062680227472350\cdot 10^{-7}\, t^{-8}-2.066685864343392\cdot 10^{-8}\, t^{-9}-4.142873131848698\cdot 10^{-9}\, t^{-10}
\nonumber\\&
-8.431737028063335\cdot 10^{-10}\, t^{-11}-1.767323272155583\cdot 10^{-10}\, t^{-12}-3.793083668117069\cdot 10^{-11}\, t^{-13}
\nonumber\\&
-8.262059820703867\cdot 10^{-12}\, t^{-14}-1.824935364466147\cdot 10^{-12}\, t^{-15}-4.085060964032317\cdot 10^{-13}\, t^{-16}
\nonumber\\&
-9.245478452546614\cdot 10^{-14}\, t^{-17}-2.112454977166614\cdot 10^{-14}\, t^{-18},\nonumber
\displaybreak[0]\\
\nonumber\\
 \Delta^+ / J_{\rm{e}}=& \, 6 -0.5 \,t^{-1}-0.05 \, t^{-2}-1.510416666666667\cdot 10^{-2} \, t^{-3}-2.042258522727273\cdot 10^{-3} \, t^{-4}
 \nonumber\\&
-2.409132234394131\cdot 10^{-4} \, t^{-5}-7.207724010454154\cdot 10^{-5} \, t^{-6} -1.073283118010885\cdot 10^{-5} \, t^{-7}
 \nonumber\\&
 -2.421548499055980 \cdot 10^{-6} \, t^{-8},\nonumber
\displaybreak[0]\\
\nonumber\\
 \Delta^- / J_{\rm{e}}=& \, 6 +0.25\, t^{-1}-0.04375 \, t^{-2}-1.223958333333333\cdot 10^{-2} \, t^{-3}-1.817215119949495\cdot 10^{-3} \, t^{-4}\nonumber\\&
-3.438872179417279\cdot 10^{-4} \, t^{-5}-8.285969963376602\cdot 10^{-5} \, t^{-6}-1.826013298799764\cdot 10^{-5} \, t^{-7}
\nonumber\\&
-3.800686519319271\cdot 10^{-6} \, t^{-8}.\nonumber
\end{align}

%
%
\section{Exact diagonalization of the effective dimer model}
%
%
This supplementary material describes the numerical results obtained for the effective  quantum dimer model (QDM) describing the string-net model [see Eq.~(6) in the manuscript] in the vicinity of $\theta=3\pi/2$. 
This effective model is a hard-core quantum dimer model on the hexagonal lattice similar to that discussed in Ref.~\cite{Moessner01} but with a different potential term.

%
%
\subsection{Finite-size clusters}
%
%

\begin{figure}[t]
\includegraphics[width=9cm]{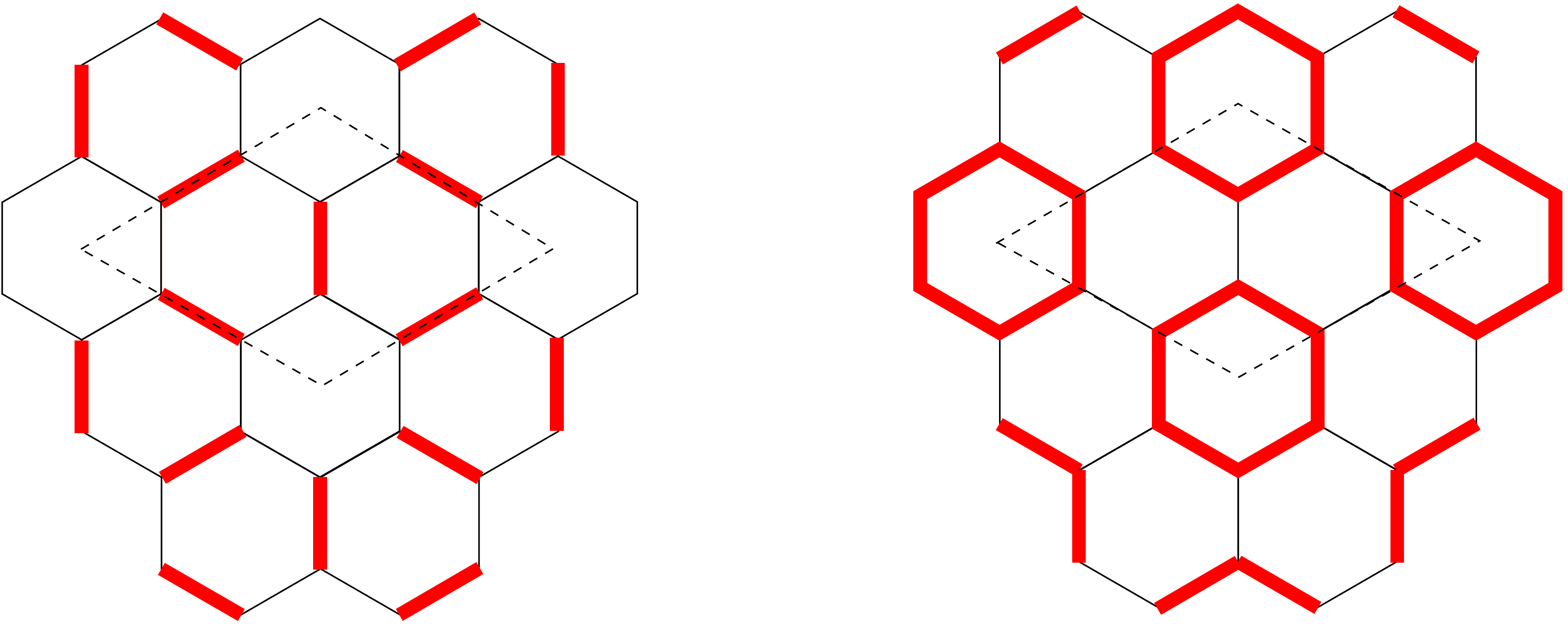}
\caption{Illustration of the star crystal (SC) (left) and the plaquette crystal (PC) (right). The SC consists in a separable state in the canonical dimer basis whereas the PC is made of resonating plaquettes with three dimers. Both crystals have the same unit cell represented by dotted lines. They also have the same momenta and point-group quantum numbers}
\label{fig:SC_PC}
\end{figure}

We performed some Lanczos diagonalizations on finite-size clusters with periodic boundary conditions up to \mbox{$N_{\rm p}=63$ plaquettes}. In this largest system, the total Hilbert space dimension is 1.648.213.392. The size and shape of the clusters we studied are given in Tab.~\ref{tab:clusters}. All these clusters are compatible with the $\sqrt{3}\times\sqrt{3}$ unit cell of the star crystal (SC) and plaquette crystal (PC) depicted in Fig.~\ref{fig:SC_PC} that turn out to be relevant for the original string-net model.

%
%
\begin{table}[h]
 \begin{tabular}{|c|c|c|}
 \hline
  $N_{\rm p}$ & $T_1$ & $T_2$ \\
  \hline
  21 & [1,4] &  \\
  27 & [3,3] &  \\
  36 & [6,0] &  \\
  39 & [2,5] &  \\
  48 & [4,-8] & \\
  54 & [6,0] & [9,-9] \\
  57 &  [-8,1] & \\
  60 & [5,5] & [6,-6] \\
  63 & [-9,3] & \\
  \hline
 \end{tabular}
 \hspace{1cm}
 \includegraphics[trim = 0cm 10cm 0cm 0cm, width=7cm]{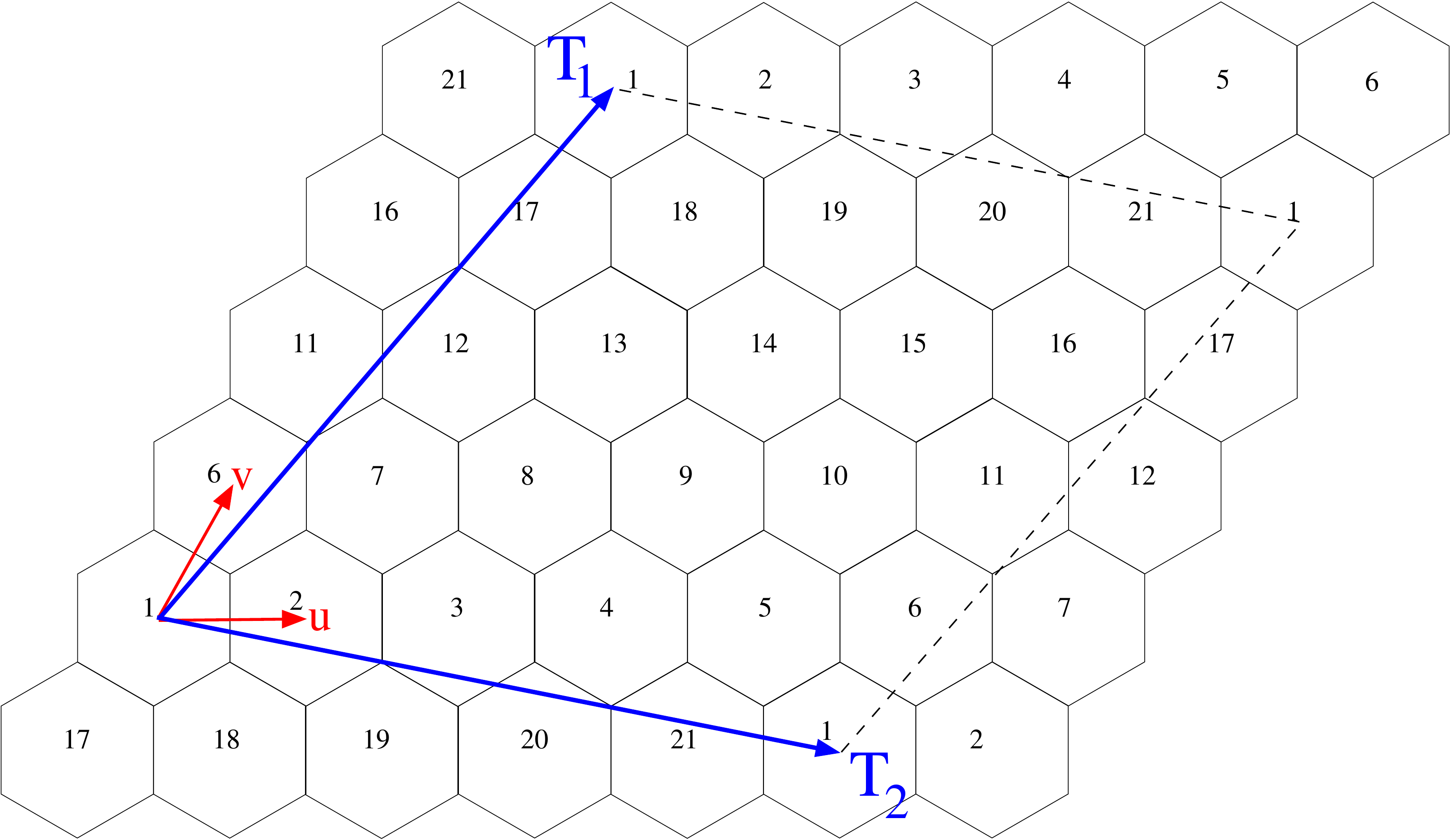}
 \vspace{0.5cm}
 \caption{Size ($N_{\rm p}$) and shape of the different clusters used in this study. Each cluster is defined by the two vectors $T_1$ and $T_2$ whose coordinates $[l_1,m_1]$ and $[l_2,m_2]$ are written in the basis $(u,v)$. When the coordinates $[l_2,m_2]$ of $T_2$ are not specified, they are defined by a $\pi/3$  rotation of $T_1$ ($l_2=l_1+m_1$ and $m_2=-l_1$) ensuring the $2\pi/3$ rotation symmetry of the cluster. Periodic boundary conditions are imposed. The cluster corresponding to $N_{\rm p}=21$ plaquettes is displayed for illustration.
}
 \label{tab:clusters}
\end{table}
%
%

%
%
%
%

%
%
\subsection{Winding sectors}
%
%
We diagonalized the Hamiltonian in each topological sector. For dimer coverings on bipartite lattices, each topological sector is labeled by two winding numbers $W_1$ and $W_2$. Here, $W_1$ ($W_2$) is defined as the ``electric flux'' crossing a loop winding around the torus in the $T_1$ ($T_2$) direction. The ``electric'' field $E_{ij}$ is defined on each bond $(ij)$ of the hexagonal lattice by the following rules: a) we assume that $i$ belongs to the even sublattice  -- if not, use $E_{ij}=-E_{ji}$; b) $E_{ij}=+1$ if $(ij)$ is occupied by a dimer; c) $E_{ij}=-1/2$ if $(ij)$ is empty (no dimer). These rules ensure that $E$ is divergence-free (each site touches one dimer and two empty bonds) and  the flux crossing a closed line is therefore invariant under local moves of this line. These winding numbers are also conserved under any local dimer move. With the definition above,  the SC and the PC found in Ref.~\cite{Moessner01} belong to the $W_1=W_2=0$ sector. Note that the staggered crystal also discussed in Ref.~\cite{Moessner01} has a macroscopic electric flux, {\it i.e.}, winding numbers proportional to the linear dimensions of the system.

%
%
\subsection{Hamiltonian}
%
%

The QDM Hamiltonian considered here is defined by
\begin{equation}
H_{\rm QDM}=-\frac{4}{J_{\rm p}} H_{\rm eff}-N_{\rm p}=\sum_p \: t\: 
\Big(
\left|\hspace{-1mm}
\begin{array}{c}\includegraphics[width=0.5cm]{./figures/flip_plaq1.pdf}\end{array} 
\hspace{-1mm}\right\rangle 
\left\langle\hspace{-1mm}
\begin{array}{c}\includegraphics[width=0.5cm]{./figures/flip_plaq2.pdf}\end{array}
\hspace{-1mm}\right|
+
\left|\hspace{-1mm}
\begin{array}{c}\includegraphics[width=0.5cm]{./figures/flip_plaq2.pdf}\end{array}
\hspace{-1mm}\right\rangle
\left\langle\hspace{-1mm} 
\begin{array}{c}\includegraphics[width=0.5cm]{./figures/flip_plaq1.pdf}\end{array}
\hspace{-1mm}\right|
\Big)
+ v\:
\left|\hspace{-1mm} 
\begin{array}{c}\includegraphics[width=0.5cm]{./figures/diag_plaq.pdf}\end{array}
\hspace{-1mm}\right\rangle 
\left\langle\hspace{-1mm} 
\begin{array}{c}\includegraphics[width=0.5cm]{./figures/diag_plaq.pdf}\end{array}
\hspace{-1mm}\right|
\label{eq:hameff}.
\end{equation}
%
%
The string-net model in the vicinity of $\theta=3\pi/2^+$ ($J_{\rm p}>0$) is described by $H_{\rm eff}$ which is related to $H_{\rm QDM}$ with $t=-1/2$ and $v=-1$ while, for $\theta=3\pi/2^-$ ($J_{\rm p}<0$), one must consider the parameters $t=1/2$ and $v=1$. When the cluster has an even number of hexagons, one can redefine the signs of dimer configurations (unitary transformation) to change $t$ into $-t$ ($v$ unchanged). However, for systems with an odd number of hexagons, the symmetry $t\to-t$ is broken. In order to compare various sizes reliably, we set $t=1/2$ in the following except when focussing on the special case $v=-1$.

%
%
\subsection{Star phase for $v\lesssim-0.5$}
%
%

\begin{figure}
\includegraphics[width=18cm]{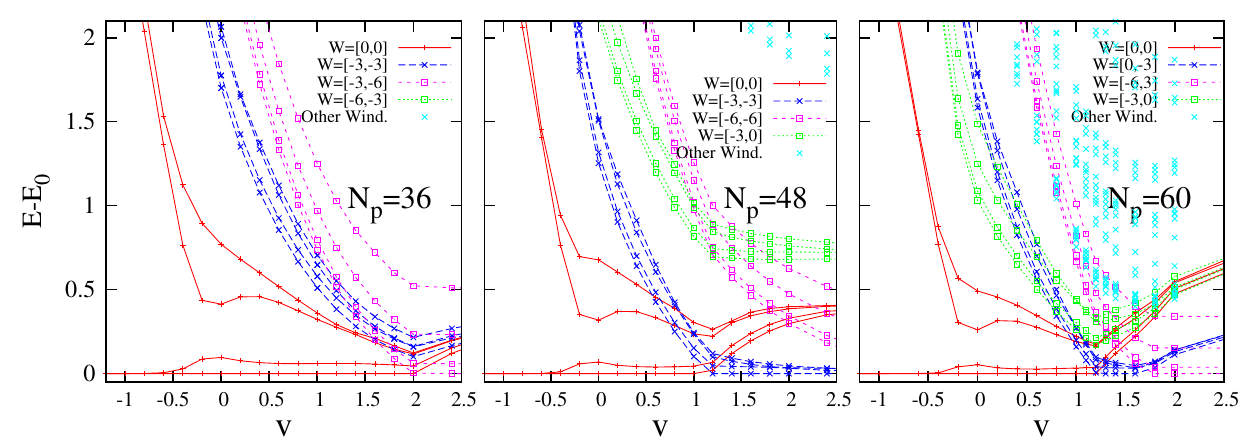}
\caption{Energy spectrum as a function of $v$ for $t=1/2$. The first four eigenvalues of each winding sector $W=[W_1,W_2]$ are shown and the ground-state energy $E_0$ is subtracted.}
\label{fig:sp}
\end{figure}

\begin{figure}
\includegraphics[width=16cm]{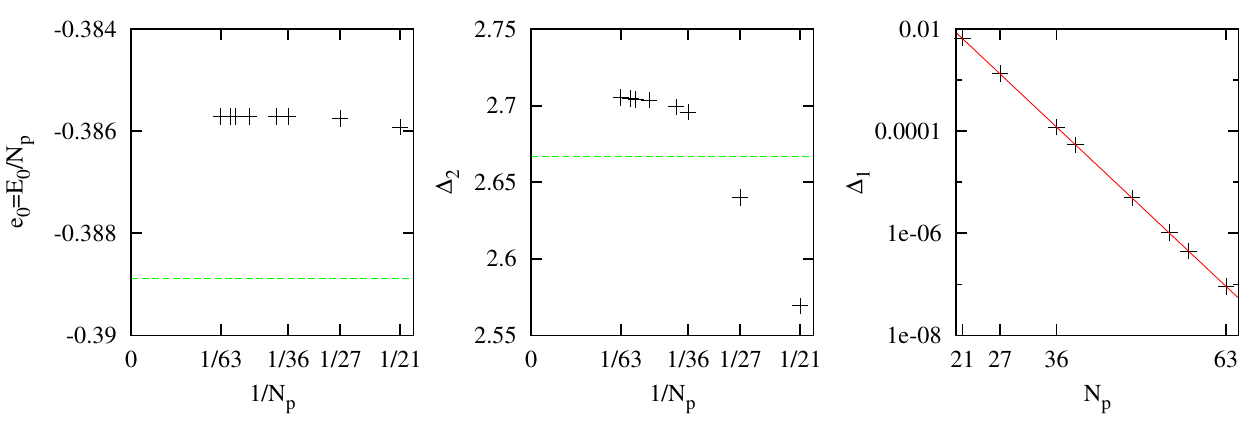}
\caption{Finite-size behavior of the ground-state energy per plaquette $e_0$ (left pannel) and the two lowest excitation energies $\Delta_2$ (middle panel) and $\Delta_1$ (log-normal plot in the right panel), for $t=-1/2$ and $v=-1$. $\Delta_2$ is finite in the thermodynamic limit  while $\Delta_1$ corresponds to a vanishingly small splitting of the ground-state multiplet (spontaneous symmetry breaking SC phase). Dotted green lines indicate the order 2 perturbative results in a $t/v$ expansion as given in Eqs.~(7-8) in the manuscript. The red line is a fit showing the exponential decrease of $\Delta_1$ with the system size. 
}
\label{fig:scaling_v-1}
\end{figure}

The evolution of the low-energy spectrum is shown in Fig.~\ref{fig:sp} for three different system sizes ($N_{\rm p}=36,48$ and 57).  For $v\lesssim-0.5$, we observe an almost perfect degeneracy of the two lowest eigenvalues, corresponding to a three-fold degenerate ground state in the thermodynamic limit (since one state belongs to a two-dimensional irreducible representation). In fact, the spectrum at $v=-1$ appears to be smoothly connected to the spectrum at $v=-\infty$  where ground states are SCs. One can check in particular that the energy splitting $\Delta_1$ of this multiplet vanishes as ${\rm e}^{-{\rm c} N_{\rm p}}$ ($c$ is a constant) at $v=-1$  when increasing the system size (right panel of Fig.~\ref{fig:scaling_v-1}). By contrast, the excitation gap $\Delta_2$ above the ground-state multiplet (middle panel of Fig.~\ref{fig:scaling_v-1}) converges to a finite value ($\Delta_2\simeq 2.71$) in the thermodynamic limit. We  note that its value  is close to the perturbative result, $\Delta_2=8/3$,  
obtained at the order 2 in a $t/v$ expansion [see Eqs.~(7-8) in the manuscript]. All these data unambiguously show that the system is indeed in a SC phase at $v=-1$, with a large excitation gap and a short correlation length.
%
%
%
%

%
%
%
%

%
%
\begin{figure}[t]
\includegraphics[width=14cm]{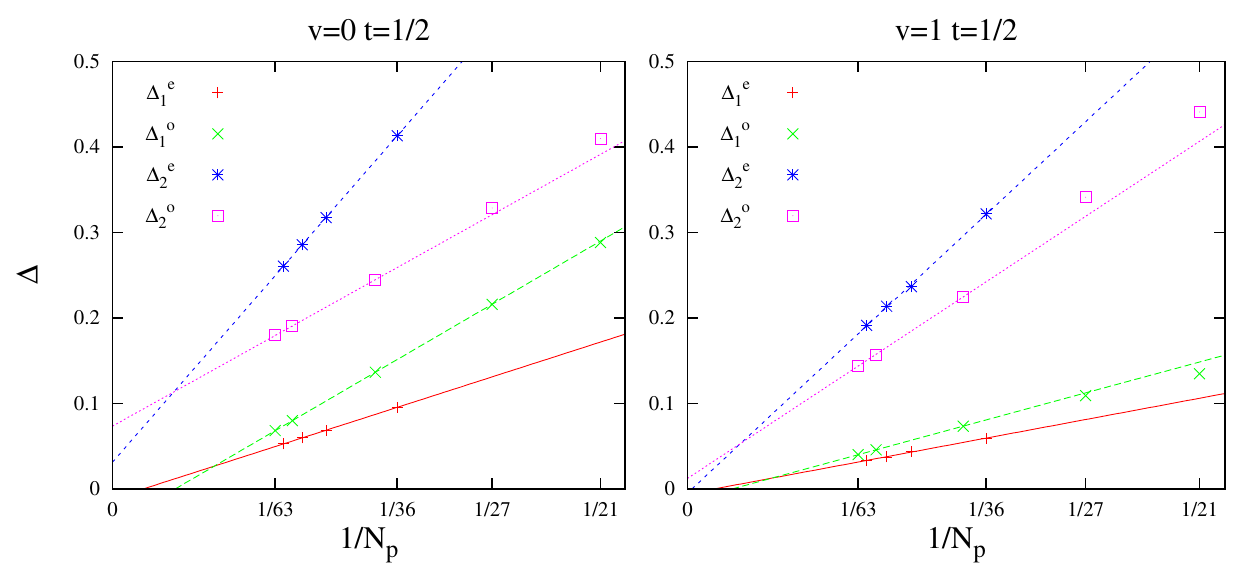}
\caption{Finite-size behavior of the two lowest excitation energies $\Delta_1$ and $\Delta_2$ at $v=0$ (left) and $v=1$ (right). Superscripts $e$ or $o$ stands for odd or even plaquette numbers. Monte-Carlo simulations \cite{Moessner01} have concluded that the system is in a PC phase at $v=0$. In such a situation the first gap $\Delta_1$ must go to zero exponentially in the thermodynamic limit (finite ground-state degeneracy due to the spontaneous symmetry breaking) while the second gap $\Delta_2$ remains finite (local excitation of the crystal). When the linear size of the system is smaller than the correlation length we have a apparent $1/N_{\rm p}$ scaling (see linear fits). Finite-size corrections are expected to become exponentially small in $N_{\rm p}$ for larger systems.  Although the available sizes do not allow to definitely conclude that the situation is the same at $v=1$, we note a rather similar behavior for $v=0$ and $v=1$.}
\label{fig:scaling_v0_v1}
\end{figure}
%
%

%
%
\subsection{Plaquette phase for $-0.5\lesssim v\lesssim1.2$}
%
%
  
At $v=0$, $H_{\rm QDM}$ coincides with the Hamiltonian studied in Ref.~\cite{Moessner01}. Using Monte Carlo simulations, it was shown that the ground state of this Hamiltonian with only the kinetic-energy term realizes a PC that has the same quantum numbers as those of the SC discussed above (same broken lattice symmetries) \cite{Moessner01}. The PC is however ``softer'', with a smaller excitation gap and a longer correlation length than in the SC phase at $v=-1$.  This can be seen from the finite-size scaling of the ground-state energy splitting ($\Delta_1$) and from the scaling of the excitation gap ($\Delta_2$) at $v=0$. The right panel of Fig.~\ref{fig:scaling_v0_v1} shows the finite-size behavior of two lowest excitation energies with an apparent $1/N_{\rm p}$ scaling indicating that the available system sizes are only comparable or smaller than the correlation length of the crystal. We also note that the transition between the SC phase to the PC phase (presumably occuring in the interval $[-0.5,0[$) is accompanied by a sudden increase in the finite-size splitting $\Delta_1$ of the ground-state multiplet (see Fig.~\ref{fig:sp}). This is also naturally explained by larger quantum fluctuations and a larger correlation length in the PC phase.

As far as the string-net model is concerned, we are interested in the point $v=1$. The spectra show a smooth evolution from $v=0$ (where the plaquette long-range order was found on large systems \cite{Moessner01}) to $v=1$. The spectra in the zero-winding sector are quantitatively very similar at $v=0$ and $v=1$. {\it From these observations we argue that the model is also likely in the same PC phase at $v=1$}. However, we  note that with the system sizes available in exact diagonalization, the scaling of the gaps  does not allow to safely conclude that the excitation gap is finite at this point (see left panel of Fig.~\ref{fig:scaling_v0_v1}). In addition, the apparent $1/N_{\rm p}$ scaling should likely cross over to an exponential behavior on larger system once the correlation length is attained. An alternative scenario would be a critical gapless  state. This is very unlikely since all QDM studied so far in two dimensions are gapped for generic values of the parameters and no mechanism producing an extended 
gapless {\it phase} in these models (as opposed to critical {\it points} such as the Rohsar-Kivelson point) is known.

%
%
\subsection{Phase for $1.2 \lesssim v$}
%
%

\begin{figure}[h]
\includegraphics[width=8cm]{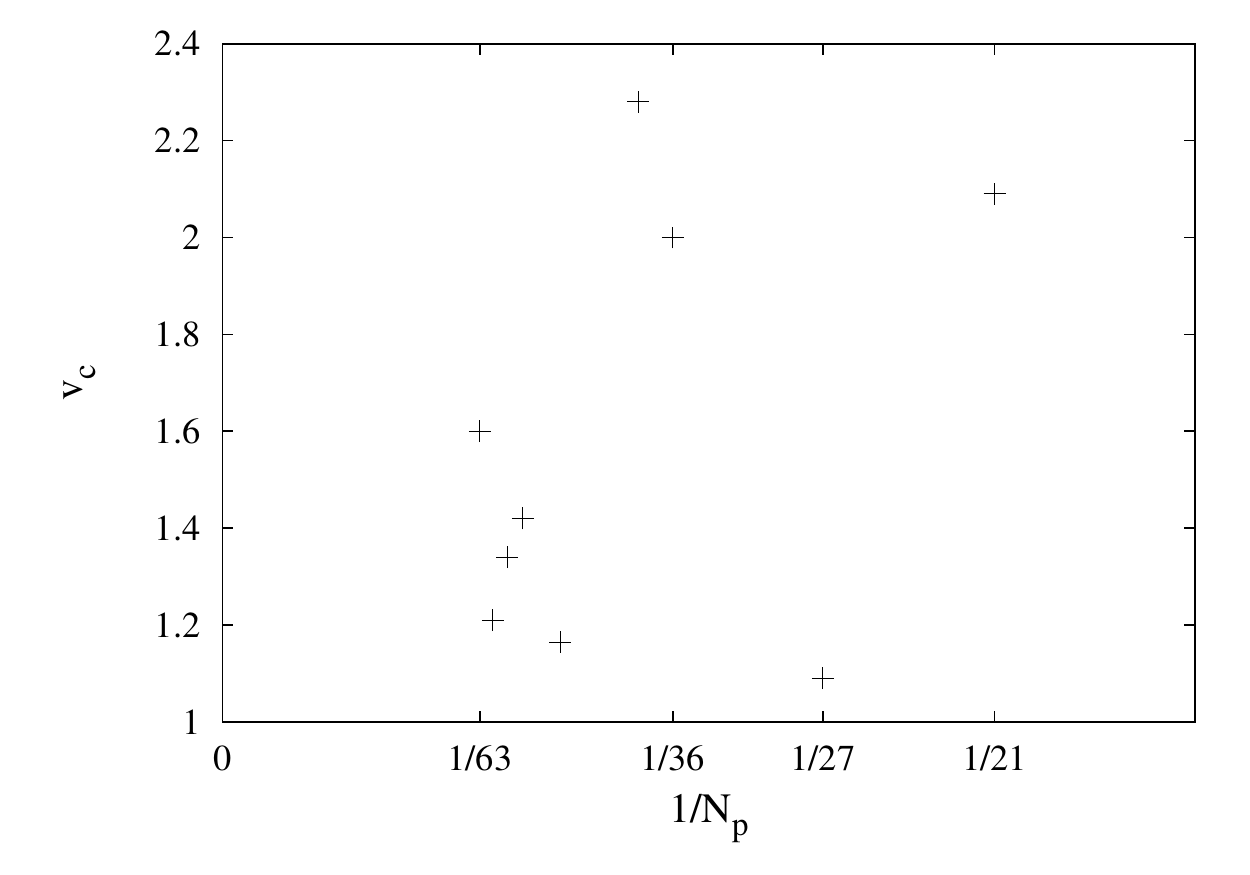}
\caption{Value of $v$ corresponding to the level crossing between the [0,0] winding sector ($v<v_{\rm c}$) and other sectors ($v>v_{\rm c}$).}
\label{fig:vc}
\end{figure}

Increasing $v$ beyond $v=1$ leads to a (first-order) transition towards a phase where the ground state is no longer in the zero-winding sector. The position $v_{\rm c}$ of this level crossing is shown in Fig.~\ref{fig:vc}, where the largest system sizes indicate that the transition takes place around $v_{\rm c}\simeq1.2$. This phase is not of direct interest for the original string-net model and will be analyzed in a future work.
%
%
%
%

\end{document}